\documentclass[11pt,letterpaper,english]{article}
\usepackage{mathptmx}
\usepackage{pdflscape}
\usepackage{iftex}
\usepackage{longtable}
\usepackage{booktabs}
\usepackage{array}
\usepackage{ragged2e}

\newcolumntype{P}[1]{>{\RaggedRight\arraybackslash}p{#1}}
\newcolumntype{C}[1]{>{\Centering\arraybackslash}p{#1}}

\usepackage{microtype}
\usepackage{float}
\usepackage{amsmath} 
\usepackage{amsthm}
\usepackage{mathrsfs}
\usepackage{bm}
\usepackage{bbm}
\usepackage{graphicx} 
\usepackage{caption}
\usepackage{setspace}
\usepackage{amssymb}
\usepackage{multirow}
\usepackage{rotating}
\usepackage{array}
\usepackage{booktabs}
\usepackage{multibib}
\usepackage{xr}
\usepackage{algorithm}
\usepackage{algpseudocode}
\usepackage{color}
\usepackage[normalem]{ulem}
\usepackage{pifont}
\usepackage{threeparttable}
\usepackage{booktabs}
\usepackage{tabularx}  
\usepackage[margin=1in]{geometry}  

\usepackage[small,compact]{titlesec}

\DeclareMathAlphabet{\mathcalligra}{T1}{calligra}{m}{n}
\usepackage[sectionbib,round]{natbib}
\usepackage{times}
\usepackage{abstract}

\usepackage[hang,flushmargin]{footmisc}
\usepackage{fullpage} 
\usepackage{hyperref}
\hypersetup{hidelinks}
\usepackage[section]{placeins}
\usepackage{footnote}
\usepackage{placeins}
\usepackage[toc,page]{appendix}
\usepackage{subcaption}
\usepackage{blkarray}
\usepackage{stackengine}
\usepackage{accents}
\usepackage{enumitem}
\usepackage[mathscr]{euscript}

\usepackage{xcolor}
\usepackage[toc,page]{appendix}
\usepackage{listings}
\bibpunct[, ]{(}{)}{;}{a}{}{,}

\theoremstyle{definition}

\theoremstyle{definition}

\newcounter{trafficTableNumber}

\usepackage[utf8]{inputenc}
\usepackage{longtable}
\usepackage{epstopdf}
\usepackage{bbm}
\usepackage{dsfont}
\usepackage{comment}
\usepackage{pgfplots}

\usepackage{pifont}
\newcommand{\cmark}{\ding{51}}

\usepackage{caption}
\theoremstyle{plain}

\newcommand{\squishlist}{
   \begin{list}{$\bullet$}
    { \setlength{\itemsep}{0pt} \setlength{\parsep}{1pt}
      \setlength{\topsep}{1pt} \setlength{\partopsep}{1pt}
      \setlength{\leftmargin}{1.5em} \setlength{\labelwidth}{1em}
      \setlength{\labelsep}{0.5em} } }

\newcommand{\squishlisttwo}{
   \begin{list}{$\bullet$}
    { \setlength{\itemsep}{0pt} \setlength{\parsep}{0pt}
      \setlength{\topsep}{0pt} \setlength{\partopsep}{0pt}
      \setlength{\leftmargin}{1em} \setlength{\labelwidth}{1.5em}
      \setlength{\labelsep}{0.5em} } }

\newcommand{\squishend}{
    \end{list}  }

\begin{document}

\title{Do Third-Party Web Traffic Estimates Preserve Causal Variation?}
\author{
Mehrzad Khosravi \thanks{Please address all correspondence to: mehrzad@uw.edu and hemay@uw.edu.}\\ University of Washington
\and
Hema Yoganarasimhan\\
University of Washington
}
\date{\today}
\maketitle

\begin{abstract}
\begin{singlespace}
Researchers increasingly rely on third-party platforms such as Similarweb and Semrush to measure web traffic when first-party analytics are unavailable. Yet these platforms report model-generated estimates rather than raw data, raising questions about whether their measures preserve the temporal and cross-source variation required for causal inference. As a motivating diagnostic, we examine reported referral traffic around two documented search-engine outages; the absence of visible discontinuities illustrates why preservation of identifying variation cannot be taken for granted. We then characterize three mechanisms---within-source smoothing, cross-source leakage, and treatment-induced calibration error---through which platform processing can generate nonclassical outcome measurement error. Analytical results and a stylized difference-in-differences simulation show that this error can attenuate, amplify, or reverse estimated treatment effects. Our findings caution against using third-party traffic measures based on black-box proprietary models for causal inference.
\end{singlespace}
\end{abstract}

\newpage

\stepcounter{table}
\setcounter{trafficTableNumber}{\value{table}}

\section{Introduction}
\label{sec:introduction}

Researchers increasingly use online traffic data to study demand, competition, media consumption, and platform regulation. In many settings, direct access to first-party server logs or analytics accounts is infeasible, leading researchers to rely on third-party measurement platforms such as Semrush and Similarweb, which provide scalable coverage of web traffic and referrals \citep{similarweb_methodology,semrush_traffic_intelligence}. In Table~\ref{tab:traffic_peer_reviewed}, we provide a non-exhaustive list of recent papers---both published articles and working papers---that rely on data from these third-party platforms. These papers span multiple disciplines, including marketing, strategy, information systems, accounting and finance, entrepreneurship, and economics, and employ empirical approaches ranging from descriptive analyses to panel regressions, event studies, difference-in-differences, and synthetic-control designs. The breadth of these applications reflects a broader shift toward data-rich empirical work: such data can expand the scope of inquiry and enable new research designs, but the underlying measurement processes are often opaque, evolving, and optimized for purposes other than scientific inference.

A key feature of the data provided by these third-party platforms is that they are \emph{model-based} estimates rather than raw counts. Similarweb describes a pipeline that combines contributory-panel data, partnerships, public data extraction, and predictive modeling \citep{similarweb_methodology}, while Semrush describes a workflow that processes panel clickstream data using neural-network prediction, anomaly filtering, and normalization \citep{semrush_data_metrics,semrush_traffic_intelligence}. These design choices are sensible for managerial benchmarking, where stability and broad comparability are valuable, but they may have adverse consequences for causal research, which often relies on sharp temporal changes or contrasts across traffic sources. Existing validation evidence documents systematic discrepancies between first- and third-party measures: using data from 86 websites, \citet{jansen2022measuring} find statistically significant differences between Google Analytics and Similarweb in total visits, unique visitors, bounce rate, and session duration. Yet it remains unclear whether---and, if so, how---these discrepancies affect causal inference analysis based on third-party traffic data.\footnote{Recent work by \citet{zhao2025strategic} finds that treatment effects estimated from Semrush and Similarweb lead to different findings compared to those based on data from raw comscore panel datasets, in a news publisher study.}

\begin{figure}[htp!]
  \centering
  \includegraphics[width=1.0\linewidth]{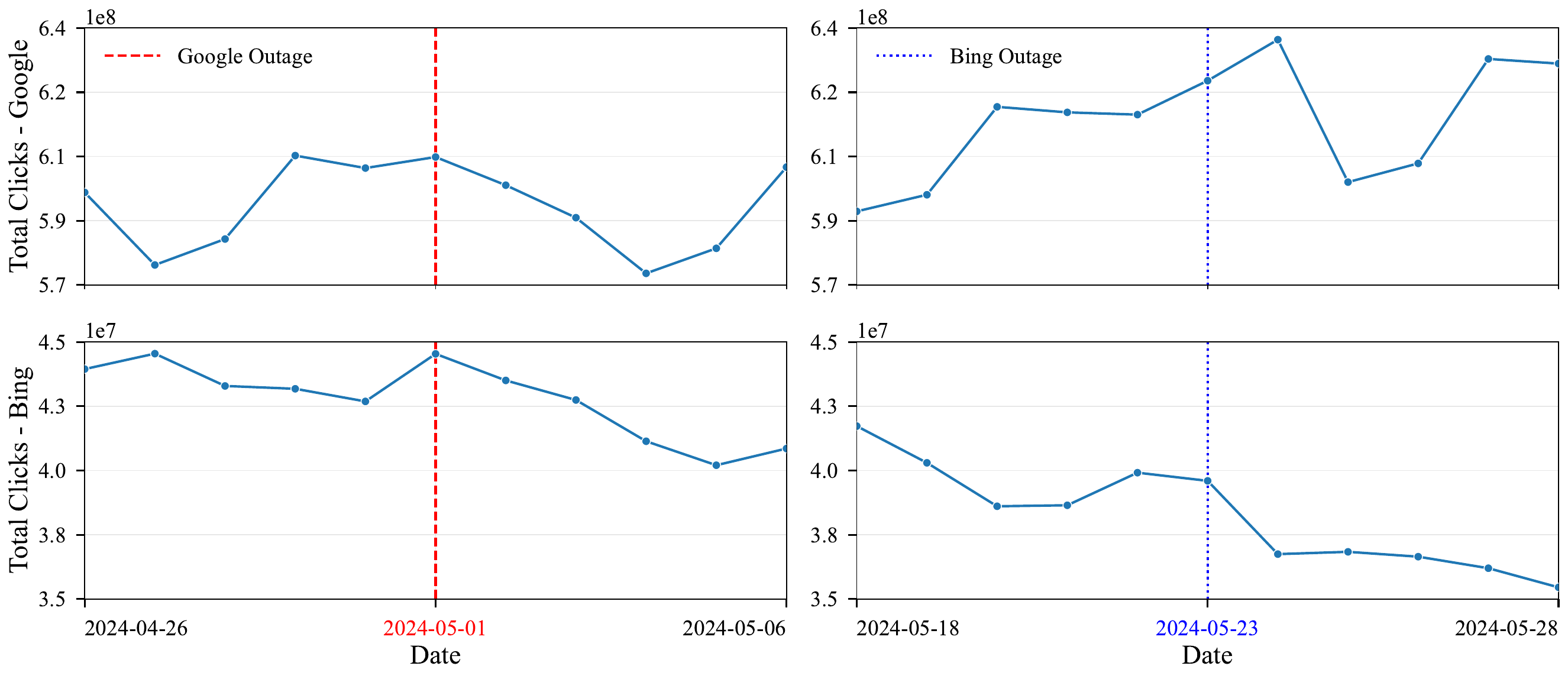}
  \caption{Daily organic search referral clicks to 7{,}791 high-traffic websites, by search engine (Dewey Data integration with Similarweb). Vertical lines mark a global Google disruption on May 1, 2024, and a multi-hour Bing outage on May 23, 2024.}
  \label{fig:google_bing}
\end{figure}

To provide an initial diagnostic of whether vendor-reported measures preserve the variation relevant for empirical inference, we examine two known, short-lived disruptions to search-engine availability. Figure~\ref{fig:google_bing} plots daily organic-search clicks from Google and Bing to 7{,}791 high-traffic domains in Similarweb data accessed through the Dewey platform. Two vertical lines mark widely reported outages: a global disruption of \texttt{google.com} on May 1, 2024 \citep{thousandeyes_google_outage_2024} and a multi-hour \texttt{bing.com} outage on May 23, 2024 \citep{techcrunch_bing_outage_2024}. Under minimal behavioral assumptions, each outage should reduce referrals from the affected search engine and may redirect some traffic toward the other engine. Yet the reported series shows no unambiguous same-day response to either event. Although this evidence is descriptive, it nevertheless suggests that exogenous events that would ordinarily be expected to leave a visible same-day imprint on raw referral counts do not produce one in the reported series.

The outage example highlights a broader inferential issue that we examine in this paper: platform-generated traffic measures may not preserve the temporal and cross-source variation on which causal research designs rely. In Section~\ref{sec:implications}, we develop a general measurement-error framework showing that the direction and magnitude of bias depend on how platform-induced measurement error covaries with the variation used for identification. We then focus on the two mechanisms most directly connected to the outage evidence. \emph{Within-source smoothing} uses a traffic source's own historical information to stabilize its reported series, whereas \emph{cross-source leakage} incorporates information from other traffic sources into a source-specific estimate. In Section~\ref{sec:simulation}, we illustrate the consequences of these mechanisms in a stylized difference-in-differences design. In Web Appendix~\ref{appsec:treatment_induced_calibration}, we extend the framework to include another source, \emph{treatment-induced calibration error}, in which treatment changes the relationship between the platform's partial traffic signals and total traffic.

We show that platform-induced measurement error need not behave like classical noise: treating model-based estimates as if they were raw observations can bias causal estimates. Within-source smoothing can suppress or otherwise reshape the sharp temporal movements used to identify treatment effects, while cross-source leakage can contaminate source-specific contrasts by transmitting the response of one traffic source into the reported outcome of another. Treatment-induced calibration error can introduce bias at an earlier stage, when treatment changes the relationship between the platform's partial traffic signals and total traffic, even in the absence of subsequent smoothing or pooling. Because these mechanisms determine how measurement error covaries with the identifying variation, estimated effects can be attenuated, amplified in absolute magnitude, or reversed in sign. The direction and magnitude of the bias depend jointly on the underlying behavioral response, the data available to the platform, and its proprietary modeling process. Further, because researchers typically observe neither the platform's underlying input data nor the transformations used to construct its reported measures, they generally cannot determine the direction or magnitude of the resulting bias in a particular application. As such, we caution researchers against using third-party traffic data for causal inference; and suggest that they are better suited to benchmarking and descriptive analysis.

\section{Why Preprocessing Can Bias Causal Inference}
\label{sec:implications}

In this section, we use measurement-error econometrics to show that platform-side data modeling can introduce nonclassical measurement errors in outcomes via two complementary mechanisms: \emph{within-source smoothing} and \emph{cross-source leakage}. Both mechanisms are stylized versions of the more complicated procedures platforms use before reporting data on their dashboards. 

To make the inferential problem concrete, consider what a researcher might seek to learn from the two outages in Figure~\ref{fig:google_bing}. Using each outage as an exogenous disruption to one search engine, the researcher could ask two questions: How much did referral traffic from the affected search engine decline, and how much of that traffic, if any, shifted to the competing search engine? We refer to the first quantity as the own-engine response and the second as the cross-engine substitution response.

Let $y_{i,s,t}$ denote the \emph{true} number of organic-referral clicks from search engine $s \in \{G,B\}$ (Google, Bing) to website $i$ on day $t$, and let $O_{r,t}$ be an indicator for an outage of engine $r$ on day $t$.\footnote{In the example, $O_{G,t}=1$ on May~1,~2024 and $O_{B,t}=1$ on May~23,~2024.} Using the outages as a running example, a convenient decomposition of these behavioral responses is
\begin{equation}
  y_{i,s,t}
  = y^{0}_{i,s,t}
  + \sum_{r \in \{G,B\}} \theta_{s,r}\,O_{r,t}
  + \varepsilon_{i,s,t},
  \label{eq:true_outage}
\end{equation}
where $y^{0}_{i,s,t}$ is the counterfactual conditional mean of true referrals in the absence of any interventions and $\varepsilon_{i,s,t}$ captures idiosyncratic deviations around that counterfactual. The coefficients $\theta_{s,r}$ summarize how an outage of engine $r$ shifts referrals from engine $s$. Under the outage logic, we expect $\theta_{G,G}<0$ and $\theta_{B,B}<0$ for the own-engine responses, while $\theta_{B,G}>0$ and $\theta_{G,B}>0$ if users substitute toward the other engine.

Researchers, however, do not observe $y_{i,s,t}$, and the third-party platform generally does not observe it either. Instead, the platform observes partial signals from sources such as user panels, data partners, first-party measurements shared by some websites, and public data. Let $\mathbf{m}_{i,r,t}$ denote the current and historical platform-observed signals associated with website $i$ and traffic source $q$ that are available when the platform constructs its day-$t$ estimate. The platform combines these inputs using proprietary models to produce the reported outcome $\widehat{y}_{i,s,t}$:
\begin{equation}
  \widehat{y}_{i,s,t}
  = f_{s}\!\left(
    \{ \mathbf{m}_{i,q,t} \}_{r \in \{G,B\}}
  \right).
  \label{eq:measurement_map}
\end{equation}
The reported outcome, therefore, need not be a classically noisy measure of true traffic. We define the platform-induced measurement error as $
  u_{i,s,t} \equiv \widehat{y}_{i,s,t}-y_{i,s,t}$. 

Although Equations~\eqref{eq:true_outage} and~\eqref{eq:measurement_map} use the search-engine outages as a running example, the underlying econometric problem applies more generally whenever third-party traffic measures are used as outcomes in causal research. In applications with a single treatment and a single traffic outcome, the source indices can be suppressed: $O_{i,t}$ can denote unit $i$'s exposure at time $t$ to a policy change, platform intervention, product launch, or other treatment; $y^0_{i,t}$ can denote the counterfactual conditional mean of true traffic in the absence of treatment; and $\theta$ can denote the causal effect of interest. When researchers study multiple traffic outcomes or channels, $s$ can continue to index those outcomes. When they study multiple interventions, $r$ can index those interventions, so that treatment exposure is represented by $O_{i,r,t}$. In the outage application, both disruptions affect all destination websites at the same time, so the unit index is suppressed and treatment exposure is represented by $O_{r,t}$. Event studies, difference-in-differences, synthetic-control methods, and related designs differ in how they construct the counterfactual and isolate the identifying variation, but the same measurement problem applies to each: the reported outcome must preserve the treatment-induced change in true traffic. To isolate distortions introduced by subsequent data processing, our main-text analysis holds fixed the relationship between true traffic and the platform's partial signals. In Web Appendix~\ref{appsec:treatment_induced_calibration}, we examine an additional mechanism in which treatment itself changes this relationship. 


\subsection{Bias Formulation}
\label{ssec:direction_of_bias}

To derive the bias from using platform-reported outcomes, we begin with the following model for true traffic:
\begin{equation}
  y_{i,s,t}
  =
  \theta_{s,r}O_{i,r,t}
  +
  \mathbf{W}_{i,r,t}'\boldsymbol{\gamma}_{s,r}
  +
  \varepsilon_{i,s,t},
  \label{eq:true_outage_regression}
\end{equation}
where $\theta_{s,r}$ is the causal effect of intervention $r$ on traffic outcome $s$, and $\mathbf{W}_{i,r,t}$ contains the controls, fixed effects, or other variables used to construct the counterfactual. 


Let $X_{i,r,t}$ denote the residual from projecting $O_{i,r,t}$ on $\mathbf{W}_{i,r,t}$. By the Frisch--Waugh--Lovell theorem, $X_{i,r,t}$ is the variation in treatment exposure that identifies $\theta_{s,r}$ \citep{frisch1933partial, lovell1963seasonal, davidson1993estimation}. If true traffic were observed, consistency would require $  \operatorname{Cov}
  \bigl(
    X_{i,r,t},
    \varepsilon_{i,s,t}
  \bigr)
  =0$. 
This is the usual behavioral identification condition: after accounting for the variables used to construct the counterfactual, treatment exposure must be unrelated to the remaining unobserved determinants of true traffic.

Researchers instead estimate the model using the platform-reported outcome
$\widehat{y}_{i,s,t}=y_{i,s,t}+u_{i,s,t}$, where $u_{i,s,t}$ is the platform-induced measurement error. The corresponding estimating equation is
\begin{equation}
  \widehat{y}_{i,s,t}
  =
  \theta_{s,r}O_{i,r,t}
  +
  \mathbf{W}_{i,r,t}'\boldsymbol{\gamma}_{s,r}
  +
  \bigl(
    \varepsilon_{i,s,t}
    +
    u_{i,s,t}
  \bigr).
  \label{eq:reported_outage_regression}
\end{equation}
Thus, even when the behavioral identification condition holds, platform-induced measurement error enters the regression's composite error term. Applying the Frisch--Waugh--Lovell theorem, the probability limit of the coefficient estimated using reported traffic is 
\begin{equation}
  \theta^{\mathrm R}_{s,r}
  \equiv
  \operatorname{plim}
  \widehat{\theta}^{\mathrm R}_{s,r}
  =
  \theta_{s,r}
  +
  \frac{
    \operatorname{Cov}
    \bigl(
      X_{i,r,t},
      u_{i,s,t}
    \bigr)
  }{
    \operatorname{Var}
    \bigl(
      X_{i,r,t}
    \bigr)
  }.
  \label{eq:general_nonclassical_bias}
\end{equation}
Equation~\eqref{eq:general_nonclassical_bias} suggests that even if treatment exposure is exogenous with respect to true traffic, the reported-outcome coefficient ($\theta^{\mathrm R}_{s,r}$, where the superscript $\mathrm R$ stands for ``Reported'') is biased whenever platform-induced measurement error covaries with the identifying variation. When
$\operatorname{Cov}(X_{i,r,t},u_{i,s,t})=0$, measurement error does not create asymptotic coefficient bias, although it may reduce precision. When this covariance is nonzero, its sign and magnitude determine the resulting distortion. For a negative true effect, a positive covariance attenuates the estimated decline and may reverse its sign, whereas a negative covariance amplifies the decline. For a positive true effect, a negative covariance attenuates the estimated increase and may reverse its sign, whereas a positive covariance amplifies the increase. Because researchers generally observe neither true traffic nor the platform-induced measurement error, they cannot estimate the covariance term in Equation~\eqref{eq:general_nonclassical_bias} from the reported series alone. 

The remainder of this section develops two stylized platform-side processes that can generate such covariance: within-source smoothing and cross-source leakage.

\subsection{Mechanism 1: Within-Source Smoothing}
\label{ssec:within_source}

Third-party platforms often smooth their traffic estimates to reduce short-run noise, limit the influence of outliers, or stabilize estimates when observations are sparse. For example, a platform may draw on the same traffic series' historical signals and predicted patterns, placing less weight on current observations when they depart sharply from the expected trajectory. We refer to this transformation as \emph{within-source smoothing}. Such smoothing can reshape causal variation whenever true traffic responds to an intervention more quickly than the platform's model adapts. In the outage application, for example, a sudden decline in referrals from a search engine may be treated as a transitory anomaly, causing the reported series to replace part of the decline with a prediction based on pre-outage traffic.\footnote{For expositional clarity, we focus on one-sided smoothing based on information available before date $t$. A platform that subsequently revises historical estimates could also use information observed after $t$, in which case the response may be redistributed both before and after the event.}

We represent within-source smoothing using the following reduced-form measurement rule:
\begin{equation}
  \widehat{y}_{i,s,t}
  =
  (1-\lambda_s)y_{i,s,t}
  +
  \lambda_s\widetilde{y}_{i,s,t},
  \qquad
  0<\lambda_s<1,
  \label{eq:smoothing}
\end{equation}
where $\widetilde{y}_{i,s,t}$ is a stabilized path constructed from the same traffic series' historical information, for example a moving average or another lag-based predictor, and $\lambda_s$ determines the weight placed on that path. Equation~\eqref{eq:smoothing} is a reduced-form representation of how smoothing transforms true traffic into the reported outcome; it does not imply that the platform directly observes $y_{i,s,t}$. Equation~\eqref{eq:smoothing} implies that platform-induced measurement error is
$u_{i,s,t}=\lambda_s(\widetilde{y}_{i,s,t}-y_{i,s,t})$.
If the stabilized path adjusts more slowly than true traffic following an intervention, the difference between them may vary systematically with the identifying variation $X_{i,r,t}$, generating $\operatorname{Cov}(X_{i,r,t},u_{i,s,t})\neq 0$.

Let $\theta^{\mathrm W}_{s,r}$ denote the population coefficient on $O_{i,r,t}$ when $\widetilde{y}_{i,s,t}$ is used as the outcome in the same regression specification used to define the true effect $\theta_{s,r}$. Because $X_{i,r,t}$ is the residualized component of $O_{i,r,t}$ that identifies the effect, the Frisch--Waugh--Lovell theorem gives
\begin{equation}
  \frac{
    \operatorname{Cov}
    \left(X_{i,r,t},u_{i,s,t}\right)
  }{
    \operatorname{Var}
    \left(X_{i,r,t}\right)
  }
  =
  \lambda_s
  \left(
    \theta^{\mathrm W}_{s,r}
    -
    \theta_{s,r}
  \right).
  \label{eq:smoothing_covariance}
\end{equation}
Substituting Equation~\eqref{eq:smoothing_covariance} into Equation~\eqref{eq:general_nonclassical_bias} yields
\begin{equation}
  \theta^{\mathrm R}_{s,r}
  =
  (1-\lambda_s)\theta_{s,r}
  +
  \lambda_s\theta^{\mathrm W}_{s,r},
  \label{eq:smoothing_response}
\end{equation}
where $\theta^{\mathrm R}_{s,r}$ is the effect obtained from the platform-reported outcome, $\hat{y}_{i,s,t}$.
This reported effect is a weighted average of the true treatment effect and the response contained in the stabilized path. 

Consider the immediate response to a search-engine outage: true referrals from the affected engine fall, while a path based on pre-outage history remains closer to the no-outage level. Reported traffic therefore falls by less than true traffic, causing measurement error to increase with outage exposure. Formally, if the stabilized path has little or no response under the same regression specification, so that $\theta^{\mathrm W}_{s,r}\approx0$, Equation~\eqref{eq:smoothing_covariance} implies
\[
  \operatorname{Cov}(X_{i,r,t},u_{i,s,t})
  \approx
  -\lambda_s\theta_{s,r}
  \operatorname{Var}(X_{i,r,t}).
\]
For a negative true effect, the covariance is positive whenever the stabilized path's response is less negative than the true response, so that $\theta^{\mathrm W}_{s,r}>\theta_{s,r}$. When the stabilized response is negligible relative to the true effect, $\theta^{\mathrm R}_{s,r}\approx(1-\lambda_s)\theta_{s,r}$, and the estimated decline is attenuated toward zero.

For a persistent intervention, a conventional trailing-window smoother gradually incorporates post-treatment observations. As the stabilized path's treatment response approaches the true response, the treatment-induced component of measurement error shrinks. Smoothing can therefore attenuate effects estimated over a short post-treatment window and create the appearance of gradual behavioral adjustment even when the true response is immediate.

\subsection{Mechanism 2: Cross-Source Leakage}
\label{ssec:cross_source}

Within-source smoothing operates across time: the platform uses a traffic series' own historical information when constructing its current reported value. A distinct concern arises when the reported estimate for one traffic source incorporates information from other sources. For example, when estimating referrals from DuckDuckGo to a website, a platform may use information associated with Google or Bing because the DuckDuckGo series is sparsely observed or because the platform estimates several related series jointly. We refer to this type of information borrowing as \emph{cross-source leakage}.

Cross-source information may improve predictive accuracy when traffic sources usually move together. However, it can become problematic for causal inference when an intervention causes those sources to respond differently. A search-engine outage is a natural example: referrals from the affected engine should decline, whereas referrals from a competing engine may increase if users substitute toward it. If the reported measure for the competing engine incorporates information from the affected engine, the latter's decline can enter the reported outcome used to measure substitution. Measurement error can consequently vary systematically with the intervention.

To represent this mechanism, we retain the reduced-form measurement structure in Equation~\eqref{eq:smoothing} but change the information used to construct the modeled component. In the previous subsection, $\widetilde{y}_{i,s,t}$ was based on source $s$'s own history. Here, we construct it using deviations in traffic from other sources:
\begin{equation}
  \widetilde{y}_{i,s,t}
  =
  y^0_{i,s,t}
  +
  \sum_{q\neq s}
  b_{s,q}
  \left(
    y_{i,q,t}-y^0_{i,q,t}
  \right),
  \qquad
  \text{with some } b_{s,q}\neq0,
  \label{eq:cross_pooling}
\end{equation}
where $q$ indexes a traffic source other than the outcome source $s$, $y^0_{i,q,t}$ denotes source $q$'s counterfactual conditional mean in the absence of the modeled interventions, and $b_{s,q}$ is the loading that determines how deviations in source $q$ enter the modeled component for source $s$.\footnote{Lagged cross-source information, shared latent factors, or other joint predictors can transmit treatment responses in a similar way. Their contribution depends on the response contained in those inputs under the researcher's regression specification.} The restriction $q\neq s$ isolates cross-source information borrowing from the within-source channel. The loadings are held constant across websites and dates in this illustration; they need not be nonnegative or sum to one. A positive loading translates to an increase in source $q$ into a higher modeled value for source $s$, whereas a negative loading translates the two sources' deviations in opposite directions.

Assume that the counterfactual conditional means for the focal and imported sources are uncorrelated with $X_{i,r,t}$, and that the behavioral identification condition in Section~\ref{ssec:direction_of_bias} holds for all these sources under the same regression specification. With $\operatorname{Var}(X_{i,r,t})>0$, the coefficient on intervention $r$ in the modeled component $\widetilde{y}_{i,s,t}$ is $
  \sum_{q\neq s}b_{s,q}\theta_{q,r}$,
where $\theta_{q,r}$ is the causal response of traffic from source $q$ to intervention $r$. Combining Equations~\eqref{eq:smoothing} and~\eqref{eq:cross_pooling} therefore gives
\begin{equation}
  \theta^{\mathrm R}_{s,r}
  =
  (1-\lambda_s)\theta_{s,r}
  +
  \lambda_s
  \sum_{q\neq s}
  b_{s,q}\theta_{q,r}.
  \label{eq:cross_pooling_response}
\end{equation}
The first term is the portion of source $s$'s true response retained in its reported outcome. The second is the response imported from the other sources used to construct the modeled component. The parameter $\lambda_s$ determines the overall importance of that component, while the loadings $b_{s,q}$ and source-specific responses $\theta_{q,r}$ determine the direction and magnitude of the imported response.

Subtracting the true effect and applying Equation~\eqref{eq:general_nonclassical_bias}, we have:
\begin{equation}
  \frac{
    \operatorname{Cov}(X_{i,r,t},u_{i,s,t})
  }{
    \operatorname{Var}(X_{i,r,t})
  }
  =
  \theta^{\mathrm R}_{s,r}-\theta_{s,r}
  =
  \lambda_s
  \left(
    \sum_{q\neq s}b_{s,q}\theta_{q,r}
    -
    \theta_{s,r}
  \right).
  \label{eq:cross_pooling_bias}
\end{equation}
Thus, measurement error covaries with the identifying variation whenever the imported response differs from the focal source's true response. The resulting bias shifts the reported effect toward the imported response.  If the imported response has the same sign as $\theta_{s,r}$ but a smaller absolute magnitude, the reported effect is attenuated. If it has the same sign but a larger absolute magnitude, the reported effect is amplified. An opposite-sign imported response offsets the retained true response and can reverse the reported effect if its weighted contribution is sufficiently large. Further, even when $\theta_{s,r}=0$, a nonzero imported response generates a spurious reported effect.

The implications are especially clear in a substitution analysis. Consider the effect of a Google outage on Bing referrals. The outcome source is Bing ($s=B$), the intervention is the Google outage ($r=G$), and the source supplying the imported information is Google ($q=G$). Suppose that Google traffic enters the modeled component for Bing with a positive loading, $b_{B,G}>0$. Equation~\eqref{eq:cross_pooling_response} becomes
\begin{equation}
  \theta^{\mathrm R}_{B,G}
  =
  (1-\lambda_B)\theta_{B,G}
  +
  \lambda_B b_{B,G}\theta_{G,G}.
  \label{eq:bing_google_leakage}
\end{equation}
If users substitute toward Bing, then $\theta_{B,G}>0$, whereas Google's own response satisfies $\theta_{G,G}<0$. True Bing referrals therefore increase while the modeled component imported from Google decreases. The reported change in Bing traffic lies below the true change, causing measurement error to decrease with outage exposure. Under the maintained assumptions, Equation~\eqref{eq:cross_pooling_bias} consequently implies $\operatorname{Cov}(X_{i,G,t},u_{i,B,t})<0$. If the reported Bing response remains positive, measured substitution is attenuated. If the imported decline exactly offsets or outweighs the retained increase, the reported response is zero or negative even though true Bing referrals increased. If there is no genuine substitution, so that $\theta_{B,G}=0$, the same mechanism generates a spurious negative Bing response. Sign reversal, therefore, does not require a negative cross-source loading: a positive loading on another source's opposite-direction response can be sufficient.

In sum, even when the intervention is exogenous and the behavioral specification is correct, cross-source leakage can cause the reported outcome to reflect responses from sources other than the one being studied. The resulting measurement error is systematically related to the identifying variation, so it can bias the estimated effect rather than merely reduce precision. 


This problem becomes harder to diagnose when within-source smoothing and cross-source leakage operate together. Their effects may reinforce or offset one another, and different combinations of true traffic responses and platform transformations can produce the same reported effect. Importantly, researchers who observe only the third-party platform's reported traffic cannot generally determine whether an estimated effect is attenuated, amplified, or reversed. Without information about the proprietary data inputs and models used by these platforms to generate the traffic estimates, the magnitude and even the sign of the effect on true traffic is thus hard to predict.

\section{Simulation}
\label{sec:simulation}

We now use stylized difference-in-differences simulations to show how the measurement mechanisms discussed above can distort an otherwise valid causal comparison. 


Consider a policy change that affects traffic to one website but leaves a second website unaffected. Researchers estimate the policy effect by comparing the change in traffic at the treated website with the change at the control website. For both websites, they observe only the traffic estimates reported by a third-party platform. We construct the simulation so that the difference-in-differences design is valid for true traffic, allowing us to isolate the consequences of using reported traffic instead.

For each website, we simulate traffic from a focal source $s$ and an auxiliary source $q\neq s$. The true traffic processes satisfy parallel trends in expectation, and the policy produces a constant effect on true focal-source traffic at the treated website. We then construct reported focal-source traffic by applying within-source smoothing and cross-source information borrowing directly to the simulated traffic paths. Across the three simulation conditions, we hold both sources' realized traffic paths and the within-source smoothing rule fixed. Only the platform's loading on the auxiliary-source series changes. Please see Web Appendix~\ref{appsec:simulation_details} for details on the simulation process and parameters.

We simulate $120$ days, indexed by $t\in\{0,\ldots,119\}$, with the policy taking effect at $t=60$. Each website therefore has $60$ pre-treatment and $60$ post-treatment observations. Let $T_i$ equal one for the treated website and zero for the control website, and let $P_t$ equal one when $t\geq60$ and zero otherwise. Their interaction, $D_{it}=T_iP_t$, indicates exposure to the policy and plays the role of $O_{i,r,t}$ in the preceding framework. Because the simulation considers a single intervention and a fixed focal source, we suppress the source and intervention indices on the treatment-effect coefficients.

Using platform-reported traffic, researchers estimate
\begin{equation}
  \widehat{y}_{i,s,t}
  =
  \alpha_i
  +
  \eta_t
  +
  \theta^{\mathrm R}D_{it}
  +
  \nu_{i,s,t},
  \label{eq:simulation_did_regression}
\end{equation}
where $\alpha_i$ and $\eta_t$ are website and day fixed effects, respectively. Let $\widehat{\theta}^{\mathrm R}$ denote the OLS estimate based on platform-reported traffic. Next, we define $\theta^{\mathrm R}\equiv\mathbb{E}[\widehat{\theta}^{\mathrm R}]$, where the expectation is over simulation draws. We also estimate the same specification using true traffic $y_{i,s,t}$ and denote the resulting coefficient estimate by $\widehat{\theta}$. By construction, $\mathbb{E}[\widehat{\theta}]=\theta$, where $\theta$ is the constant policy effect on true focal-source traffic.


To connect this comparison to Equation~\eqref{eq:general_nonclassical_bias}, let $X_{it}$ denote the sample residual from projecting $D_{it}$ on the website and day fixed effects, and let $u_{i,s,t}=\widehat{y}_{i,s,t}-y_{i,s,t}$. The two estimates satisfy the finite-sample identity
\begin{equation}
  \widehat{\theta}^{\mathrm R}-\widehat{\theta}
  =
  \frac{
    \sum_{i,t}X_{it}u_{i,s,t}
  }{
    \sum_{i,t}X_{it}^{2}
  },
  \label{eq:simulation_bias_identity}
\end{equation}
which isolates the component of measurement error aligned with the treatment variation used by the estimator. Because the true-outcome estimator is unbiased under the simulation design, averaging this difference across simulation draws gives the bias in the reported-outcome estimator. This is the finite-sample counterpart of the covariance-bias relationship developed in Section~\ref{ssec:direction_of_bias}.


We consider three simulation conditions in which the true policy effects are identical:
\squishlist
    \item \textbf{Attenuation.} The platform imports a smaller decline from the auxiliary source. Together with the delayed adjustment from within-source smoothing, this produces a reported decline smaller than the true effect.

    \item \textbf{Amplification.} The platform places a larger positive loading on the same auxiliary-source decline. The imported decline is strong enough to outweigh the attenuation from smoothing, producing a reported decline larger than the true effect.

    \item \textbf{Sign reversal.} The platform places a negative loading on the auxiliary source, translating its decline into an increase in the modeled component. This increase outweighs the declines retained from current and historical focal-source traffic, so reported traffic rises even though true traffic falls.
\squishend
In Web Appendix~\ref{appsec:simulation_details}, we  provide the measurement rules, parameter values, and implementation details for each condition.

\begin{figure}[H]
  \centering
  \includegraphics[width=\linewidth]
    {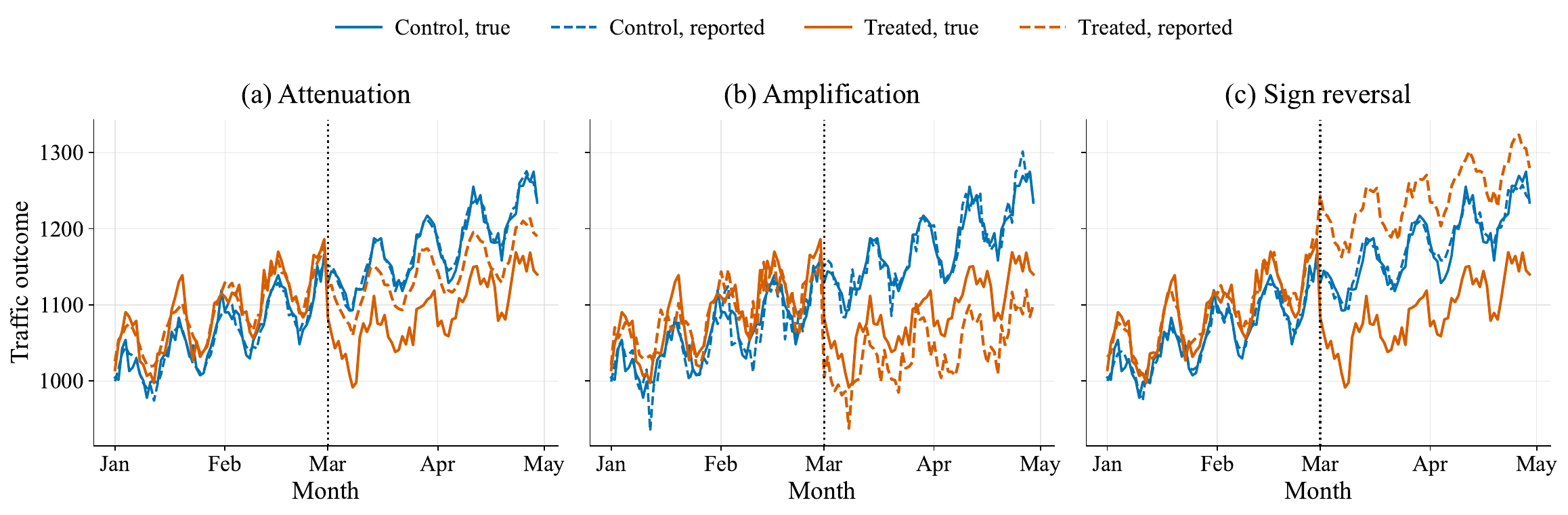}
  \caption{True and platform-reported traffic under the three simulation conditions. Each panel plots traffic for the treated and control websites. Solid lines represent true traffic, dashed lines represent platform-reported traffic, blue lines represent the control website, and orange lines represent the treated website. The vertical dotted line marks the policy change on day 60.}
  \label{fig:simulation_traffic_paths}
\end{figure}

Figure~\ref{fig:simulation_traffic_paths} presents the traffic paths generated under the three simulation conditions. The underlying true-traffic processes and realized traffic paths are identical across the panels. Differences among the platform-reported paths therefore arise entirely from changes in how the platform transforms the same underlying traffic data.

Table~\ref{tab:simulation_did_results} reports the corresponding difference-in-differences estimates for the three simulations. The true-traffic estimate is identical across the three conditions because the underlying data-generating process for the traffic data does not change. The reported-traffic estimates differ because each condition applies a different platform measurement rule to those same data.

\begin{table}[htp!]
  \centering
  \small
  \caption{Difference-in-Differences Estimates Using True and Reported Traffic}
  \label{tab:simulation_did_results}
  \begin{threeparttable}
    \begin{tabular}{lcc}
      \toprule
      Condition
      & True traffic
      & Reported traffic \\
      \midrule

      Attenuation
      & $-115.753$
      & $-68.203$ \\
      & $(3.873)$
      & $(1.558)$ \\[0.4em]

      Amplification
      & $-115.753$
      & $-170.275$ \\
      & $(3.873)$
      & $(4.625)$ \\[0.4em]

      Sign reversal
      & $-115.753$
      & $33.869$ \\
      & $(3.873)$
      & $(2.266)$ \\

      \bottomrule
    \end{tabular}

    \begin{tablenotes}[flushleft]
      \footnotesize
      \item Notes: Exact Gaussian DGP-based
      standard errors appear in parentheses; see Web Appendix~\ref{appssec:simulation_design}.
      Two-sided tests of zero expected DiD coefficients give $p<0.001$.
    \end{tablenotes}
  \end{threeparttable}
\end{table}

Taken together, these results show that a correctly specified difference-in-differences design can recover the policy effect from true traffic while producing a substantially different estimate from platform-reported traffic. Across the three conditions, the true data-generating processes, treatment assignment, and estimating equation remain unchanged; only the platform's measurement rule varies. This change alone is sufficient to attenuate, amplify, or reverse the sign of $\widehat{\theta}^{\mathrm R}$ relative to $\widehat{\theta}$. Further, we see that including website and day fixed effects does not eliminate bias when the platform's measurement process creates covariation between the platform-induced measurement error and the causally identifying variation. Without information about that process or validation against first-party traffic, researchers cannot determine the magnitude or direction of the distortion from the reported series alone.

\section{Conclusion}
\label{sec:conclusion}

Third-party web-traffic platforms provide researchers with broad and otherwise difficult-to-obtain measures/estimates of website visits, referrals, and traffic sources. However, these measures are not direct behavioral counts. Rather, they are model-generated estimates constructed from incomplete signals, auxiliary information, and proprietary transformations that researchers generally cannot observe. Although these procedures may improve stability and predictive accuracy for benchmarking purposes and broad descriptives, they can also reshape the variation that researchers use for causal inference.

This concern is especially important for research designs that rely on sharp policy changes over time or differences across traffic sources. A reported series may accurately represent average traffic levels while failing to preserve a policy change or exogenous disruption or a reallocation of traffic across sources. Within-source smoothing can distribute an abrupt response across multiple periods, while cross-source leakage can import the response of one traffic source into the reported measure of another. The resulting measurement error is nonclassical because it covaries with the variation identifying the treatment effect. Consequently, comparison groups, fixed effects, and conventional standard errors do not necessarily eliminate the resulting bias. Depending on the platform's measurement process, the estimated effect may be attenuated, amplified, or reversed in sign.

We motivate this concern in two complementary and illustrative ways. First, we examine daily organic-search referrals to 7,791 high-traffic websites around two widely documented Google and Bing outages and find no clear visible response in the platform-reported series. This provides a suggestive diagnostic: short-lived disruptions that should affect referrals are not apparent in the reported data, raising the question of whether model-generated measures preserve the variation required for causal inference. Second, we introduce within-source smoothing and cross-source leakage as stylized examples of how a platform's data-modeling process could reshape that variation. Our difference-in-differences simulation shows that changes in the platform's measurement rule alone can generate attenuation, amplification, or sign reversal while the underlying traffic processes and treatment effect remain fixed. Web Appendix~\ref{appsec:treatment_induced_calibration} develops treatment-induced calibration error as an additional mechanism that can arise when treatment changes the relationship between a platform's partial signals and total traffic. Together, these exercises illustrate the range of inferential consequences that opaque data modeling can potentially produce.

Our analysis has two main takeaways. First, model-based traffic estimates can significantly bias estimated treatment effects in causal analysis. Second, without independent, design-specific validation, third-party traffic estimates should not be relied on for causal inference and are better suited to benchmarking and descriptive analysis.


\section*{Funding and Competing Interests Declaration} Author(s) have no competing interests to declare.

\clearpage
\begingroup
\setlength{\paperwidth}{11in}
\setlength{\paperheight}{8.5in}
\ifPDFTeX
  \pdfpagewidth=11in
  \pdfpageheight=8.5in
\fi
\setlength{\textwidth}{9in}
\setlength{\textheight}{6.5in}
\setlength{\oddsidemargin}{0in}
\setlength{\evensidemargin}{0in}
\setlength{\topmargin}{0in}
\setlength{\headheight}{0pt}
\setlength{\headsep}{0pt}
\setlength{\footskip}{30pt}
\setlength{\columnwidth}{\textwidth}
\setlength{\linewidth}{\textwidth}
\setlength{\hsize}{\textwidth}
\setlength{\vsize}{\textheight}
\makeatletter
\setlength{\@colht}{\textheight}
\setlength{\@colroom}{\textheight}
\makeatother

\footnotesize
\setlength{\tabcolsep}{3pt}
\renewcommand{\arraystretch}{1.15}



\edef\tableCounterBeforeTraffic{\number\value{table}}
\setcounter{table}{\numexpr\value{trafficTableNumber}-1\relax}

\setlength{\LTcapwidth}{\textwidth}
\begin{longtable}{@{}P{5.2cm}P{7.2cm}P{7.4cm}C{2.2cm}@{}}

\caption[Papers Using Similarweb or Semrush Traffic Data]{Papers Using Similarweb or Semrush Traffic Data\newline
\textsuperscript{*}A checkmark flags potential bias from third-party data modeling when central findings remain dependent on modeled traffic after accounting for independent benchmarks and triangulation. An unmarked entry credits relevant independent benchmarking of the traffic measure or independent evidence supporting the main substantive conclusion. Evidence about related outcomes alone is insufficient if a central finding remains dependent on modeled traffic. Unmarked studies may still contain model-dependent traffic estimates; neither designation establishes the presence or absence of bias. A question mark indicates insufficient information for classification.}
\label{tab:traffic_peer_reviewed}\\

\toprule
\textbf{\ifPDFTeX\else\fi Paper} &
\textbf{Platform and type of data} &
\textbf{Main analysis} &
\textbf{Inference risk}\textsuperscript{*} \\
\midrule
\endfirsthead

\multicolumn{4}{c}%
{{\tablename\ \thetable{} -- continued from previous page}}\\
\toprule
\textbf{\ifPDFTeX\else\fi Paper} &
\textbf{Platform and type of data} &
\textbf{Analysis Method} &
\textbf{Inference risk}\textsuperscript{*} \\
\midrule
\endhead

\midrule
\multicolumn{4}{r}{\footnotesize Continued on next page}\\
\endfoot

\bottomrule
\endlastfoot

\citet{carballa_etal_2026}\newline
\emph{Strategic Management Journal}
&
\textbf{Similarweb:} directed domain-to-domain referral visits for 241 platforms operating in Europe.
&
Network analysis and qualitative triangulation.
&
\cmark
\\
\addlinespace

\citet{liu_wang_2026}\newline
\emph{World Development}
&
\textbf{Semrush:} visits, users, demographics, device use, and engagement for the 40 most visited generative-AI tools across more than 200 economies; supplemented with Google Trends.
&
Descriptive analysis and cross-country regressions.
&
\cmark
\\
\addlinespace

\citet{pape2026competition}\newline
\emph{Marketing Science}
&
\textbf{Semrush:} monthly country-level desktop/mobile traffic, channel, and engagement measures for Google Maps, Google, and Bing Maps.
&
Difference-in-differences.
&
\cmark
\\
\addlinespace

\citet{wright_2026}\newline
\emph{Organization Science}
&
\textbf{Similarweb:} country-level page visits to startup websites before and after BetaList featuring, used to distinguish local from foreign users.
&
Panel and event-time analyses.
&
\cmark
\\
\addlinespace

\citet{armstrong_konchitchki_zhang_2025}\newline
\emph{The Accounting Review}
&
\textbf{Similarweb:} monthly total visits and page views for 1,067 U.S. public firms, linked to Compustat, I/B/E/S, and CRSP.
&
Predictive regressions and portfolio analysis.
&
\cmark
\\
\addlinespace

\citet{cheng_hsu_2025}\newline
\emph{Data Technologies and Applications}
&
\textbf{Similarweb:} total visits and ChatGPT-driven traffic for 80 news websites in the United States and Taiwan.
&
Partial least squares structural equation modeling (PLS-SEM) and multi-group analysis.
&
\\
\addlinespace

\citet{dehaan_lawrence_litjens_2025}\newline
\emph{Management Science}
&
\textbf{Similarweb:} approximately 2.7 billion website visits following Google searches for S\&P 500 ticker symbols, including click destinations.
&
Classification, validation, simulations, and regressions.
&
? 
\\
\addlinespace

\citet{burtch2024consequences}\newline
\emph{Scientific Reports}
&
\textbf{Similarweb:} daily traffic for major websites, including Stack Overflow, around ChatGPT's release; linked to contribution data from Stack Overflow, Stack Exchange, and Reddit.
&
Synthetic control and difference-in-differences.
&
\cmark
\\
\addlinespace

\citet{cao_koning_nanda_2024}\newline
\emph{Management Science}
&
\textbf{Similarweb:} monthly website visits for Product Hunt ventures from six months before through twelve months after launch, used as a venture-growth outcome.
&
Triple-differences analysis.
&
\cmark
\\
\addlinespace

\citet{furstenau_etal_2023}\newline
\emph{Information Systems Research}
&
\textbf{Semrush:} monthly unique organic visitors for six large transaction platforms from 2013--2020, aggregated across country-specific domains.
&
Panel vector autoregression and Granger-causality tests.
&
\\
\addlinespace

\citet{congiu_sabatino_sapi_2022}\newline
\emph{Information Economics and Policy}
&
\textbf{Similarweb:} visits, engagement, and traffic-channel measures for a large panel of European and U.S. websites.
&
Difference-in-differences.
&
\cmark
\\
\addlinespace

\citet{dushnitsky_piva_rossilamastra_2022}\newline
\emph{Strategic Management Journal}
&
\textbf{Similarweb:} web-traffic measures used as a performance outcome for the sample of 211 crowdfunding platforms operating in EU-15 countries.
&
Clustering and performance regressions.
&
\\
\addlinespace

\citet{koning_hasan_chatterji_2022}\newline
\emph{Management Science}
&
\textbf{Similarweb:} website visits and page views for a large panel of approximately 35,000 technology startups, linked to technology-adoption and venture data.
&
Panel fixed-effects and adoption-timing analyses.
&
\cmark
\\
\addlinespace

\citet{calzada2020news}\newline
\emph{Marketing Science}
&
\textbf{Similarweb:} domain-level daily visits and engagement for European newspapers.
&
Difference-in-differences.
&
\cmark
\\
\addlinespace

\citet{cranney2026global}\newline
\emph{Working Paper}
&
\textbf{Similarweb:} estimated gender composition of users of major GenAI web versions, combined with survey and experimental evidence from many countries.
&
Evidence synthesis and pooled comparisons.
&
\\
\addlinespace

\citet{huang2026search}\newline
\emph{Working Paper}
&
\textbf{Similarweb:} daily traffic to major GenAI platforms and conventional search engines, including referral destinations and website-category information around earnings announcements.
&
Fixed-effects panel regressions.
&
\cmark
\\
\addlinespace

\citet{kim2026generative}\newline
\emph{Working Paper}
&
\textbf{Semrush:} website traffic and customer-acquisition measures for U.S. software startups, linked to technology adoption, hiring, financing, and workforce-composition.
&
Difference-in-differences.
&
\cmark
\\
\addlinespace

\citet{zhou2026llms}\newline
\emph{Working Paper}
&
\textbf{Semrush:} monthly domain-level referral estimates across 15 channels for 8,082 MBFC-rated news and media domains, including ChatGPT, Perplexity, Gemini, Claude, Grok, and traditional channels.
&
staggered difference-in-differences and event studies.
&
\cmark
\\
\addlinespace

\citet{cao2025vibecoding}\newline
\emph{Working Paper}
&
\textbf{Semrush:} post-launch website visits for Product Hunt ventures, linked to founder histories, venture funding, and employment data.
&
Difference-in-differences and Poisson fixed-effects models.
&
\cmark
\\
\addlinespace

\citet{han2025should}\newline
\emph{Working Paper}
&
\textbf{Semrush:} website traffic for news organizations entering cooperation or content-sharing arrangements with OpenAI and comparison publishers.
&
Generalized synthetic control.
&
\cmark
\\
\addlinespace

\citet{zhao2025strategic}\newline
\emph{Working Paper}
&
\textbf{Similarweb and Semrush:} Similarweb daily domain visits, Semrush traffic by channel, and Comscore browsing data, combined with historical \texttt{robots.txt}, webpage, archive, and employment data.
&
Staggered difference-in-differences.
& \cmark
\\
\addlinespace

\citet{miller2024impact}\newline
\emph{Working Paper}
&
\textbf{Similarweb:} visits, users, page impressions, and engagement for 6,387 websites in 24 industries, spanning 11 months before and 19 months after GDPR; includes an external AGOF benchmark for a subset of German sites.
&
Generalized synthetic control.
&
\\
\addlinespace

\citet{xavier2024web}\newline
\emph{Working Paper}
&
\textbf{Similarweb:} monthly visits for more than 250,000 websites.
&
Descriptive analysis of web use and concentration.
&
\cmark
\\
\addlinespace

\citet{calzada2021who}\newline
\emph{Working Paper}
&
\textbf{Similarweb:} daily desktop search referrals and total visits for 606 news outlets in 15 European countries; supplemented with Ahrefs keyword and ranking data.
&
First-difference OLS, instrumental variables, and difference-in-differences.
&
\cmark
\\
\addlinespace

\citet{cipriani2020sophisticated}\newline
\emph{Working Paper}
&
\textbf{Semrush:} estimated visits to prime money-market-fund websites, separated by institutional- and retail-oriented funds.
&
Panel fixed-effects comparisons.
&
\\
\addlinespace

\\

\end{longtable}
\setcounter{table}{\tableCounterBeforeTraffic}

\clearpage
\endgroup
\makeatletter
\vsize=\textheight
\@colht=\textheight
\@colroom=\textheight
\makeatother

\ifPDFTeX\else\fi

\bibliographystyle{abbrvnat}
\bibliography{ref}

\singlespacing

\newpage

\begin{appendices}

\setcounter{table}{0}
\setcounter{figure}{0}
\setcounter{equation}{0}
\setcounter{page}{0}
\renewcommand{\thetable}{\Alph{section}\arabic{table}}
\renewcommand{\thefigure}{\Alph{section}\arabic{figure}}
\renewcommand{\theequation}{\Alph{section}\arabic{equation}}
\renewcommand{\thepage}{\roman{page}}
\pagenumbering{roman}

\section{Treatment-Induced Calibration Error}
\label{appsec:treatment_induced_calibration}

\setcounter{equation}{0}
\setcounter{table}{0}

\subsection{Why Platforms Calibrate Partial Traffic Signals}

A third-party traffic platform faces a basic prediction problem as it does not observe a census of every visit to every website. Instead, it observes a collection of partial signals, such as activity in a contributory panel, first-party measurements shared by some websites, partner data, and public information. These are the inputs that we abstractly denote by $\mathbf{m}_{i,r,t}$ in Equation~\eqref{eq:measurement_map}. However, its clients, including brands and researchers, want estimates of total traffic. The platform must therefore translate the activity it observes into an estimate of the activity it does not observe. Similarweb describes this process in terms of weighting, calibration, and predictive modeling, while Semrush describes normalization and machine-learning prediction \citep{similarweb_methodology,semrush_traffic_intelligence}.

This calibration has a useful commercial purpose. Suppose a platform observes 500 visits to a website and concludes that its inputs represent 5 percent of the relevant population. Reporting only the 500 observed visits would not answer the client's question. The platform instead extrapolates to an estimated total of 10,000 visits. A comparable adjustment is needed for each country, device, user group, and traffic source. The adjustment is especially important when the underlying signals are sparse.

Calibration aims to improve accuracy by learning the typical relationship between observed signals and total traffic. The inferential risk arises when a treatment changes that relationship. For example, a product launch may suddenly attract younger, foreign, or mobile users. A privacy regulation may change which users can be observed. A platform redesign may send visitors through a new application or referral path. These events can alter how representative the platform's signals are while the outcome of interest changes. If the calibration model does not fully adjust, its error becomes treatment-related. We call this mechanism \emph{treatment-induced calibration error}.

\subsection{A Concrete Example}

Consider a small startup that is featured on a product-discovery platform, as in \citet{cao_koning_nanda_2024}. Before featuring, suppose the startup receives 1,000 true visits. The traffic platform effectively observes 10 percent of them, or 100 visits. It also correctly estimates its coverage to be 10 percent. The reported estimate is therefore
\begin{equation}
  \widehat y_0
  =\frac{100}{0.10}
  =1{,}000.
  \label{eq:calibration_example_pre}
\end{equation}
Thus, the pre-treatment estimate is correct.

Now suppose featuring raises true traffic from 1,000 to 1,200 visits. The new visitors may differ from the old visitors in ways that matter for measurement. Assume that they make the website easier to observe, so the platform's inputs now contain 15 percent of all visits. The platform consequently observes 180 visits. However, the model interprets the new traffic as if its coverage had risen to 20 percent rather than 15 percent. It reports
\begin{equation}
  \widehat y_1
  =\frac{180}{0.20}
  =900.
  \label{eq:calibration_example_post}
\end{equation}
True traffic increased by 200 visits, but reported traffic decreased by 100 visits. Actual coverage rose from 10 to 15 percent, while the platform estimated that it rose to 20 percent. The platform therefore treated the 180 observed visits as a larger fraction of total traffic than they actually represented. This illustrative example produces sign reversal, but smaller calibration errors could instead attenuate or amplify the true increase.

\subsection{The Measurement Map}

We next state the mechanism more generally. To keep the notation simple, we suppress the website and traffic-source subscripts, use the scalar $m_t$ to represent the relevant partial signal from the set in Equation~\eqref{eq:measurement_map}, and compare one pre-treatment period, $t=0$, with one post-treatment period, $t=1$. Let $y_t$ denote true traffic, and define $p_t\in(0,1]$ as the platform's effective coverage of true visits in period $t$. Let $\widehat p_t>0$ denote the coverage estimate used to construct reported traffic. The maintained conditional-mean assumption is
\begin{equation}
  \mathbb{E}[m_t\mid y_t,p_t,\widehat p_t]
  =p_t y_t.
  \label{eq:calibration_observed_signal}
\end{equation}
If effective coverage is 10 percent, Equation~\eqref{eq:calibration_observed_signal} says that the platform expects its inputs to contain signals corresponding to 10 percent of true visits, even after conditioning on the coverage estimate used by the platform. This proportional representation is a simplifying assumption. It holds, for example, if $\widehat p_t$ is based on information that is mean-independent of the idiosyncratic component of $m_t$ conditional on $y_t$ and $p_t$. If $\widehat p_t$ instead uses that same idiosyncratic realization, Equation~\eqref{eq:calibration_observed_signal} must be imposed directly, and the derivation below does not need to hold without it.

The platform does not observe $p_t$ directly. If it did, it could recover total traffic without calibration bias, in expectation. Instead, it uses $\widehat p_t$, constructed from its available information and modeling process. A simple inverse-coverage representation of reported traffic is
\begin{equation}
  \widehat y_t
  =\frac{m_t}{\widehat p_t}.
  \label{eq:calibration_reported_traffic}
\end{equation}
The intuition is the same as in the example. The smaller the fraction of traffic that the platform believes it observes, the more it must expand the observed signal. The platform does not estimate one scalar $\widehat p_t$ or perform this exact division. Its actual model can use many inputs and nonlinear transformations. Equation~\eqref{eq:calibration_reported_traffic} is a reduced-form representation of the net expansion performed by that richer model.

Combining Equations~\eqref{eq:calibration_observed_signal} and~\eqref{eq:calibration_reported_traffic} gives the expected reported traffic:
\begin{equation}
  \mathbb{E}[\widehat y_t\mid y_t,p_t,\widehat p_t]
  =\frac{p_t}{\widehat p_t}y_t
  =\rho_t y_t,
  \qquad
  \rho_t\equiv\frac{p_t}{\widehat p_t}.
  \label{eq:calibration_ratio}
\end{equation}
The ratio $\rho_t$ summarizes calibration accuracy. When $\rho_t=1$, estimated coverage equals effective coverage and reported traffic is correct in expectation. When $\rho_t<1$, the platform overestimates its coverage and therefore reports too little total traffic. When $\rho_t>1$, it underestimates its coverage and reports too much traffic. Importantly, $\rho_t$ is the ratio between actual and estimated coverage, not the coverage rate.

Treatment can change actual effective coverage $p_t$ by changing the users, devices, countries, or traffic paths represented in the platform's inputs. It can also change estimated coverage $\widehat p_t$ through the model's response to those inputs. We refer to a treatment-induced change in the ratio $\rho_t=p_t/\widehat p_t$ as treatment-induced calibration error. Such a change can arise when estimated coverage fails to adjust to a change in actual coverage, adjusts by the wrong amount, or changes when actual coverage does not. If actual and estimated coverage change proportionally, $\rho_t$ remains constant, so no treatment-induced calibration component arises, although a stable scaling error may remain.

\subsection{Bias in the Reported Treatment Effect}

For this two-period illustration, suppose that, in the absence of treatment, true traffic would remain at its pre-treatment level $y_0$ and the calibration ratio would remain at its pre-treatment value $\rho_0$. Under this assumption, any change from $\rho_0$ to the realized post-treatment calibration ratio $\rho_1$ is attributable to treatment. Let the true treatment effect be $\theta$, so that observed post-treatment traffic is
\begin{equation}
  y_1=y_0+\theta.
  \label{eq:calibration_true_effect}
\end{equation}
A positive $\theta$ is a true traffic increase and a negative $\theta$ is a true decline. Using Equation~\eqref{eq:calibration_ratio}, the expected before--after change in reported traffic is
\begin{align}
  \theta^{\mathrm R}
  &\equiv \rho_1y_1-\rho_0y_0 \notag\\
  &=\rho_1(y_0+\theta)-\rho_0y_0 \notag\\
  &=\underbrace{\rho_1\theta}_{\text{measured part of the true effect}}
    +\underbrace{(\rho_1-\rho_0)y_0}_{\text{baseline revaluation}}.
  \label{eq:calibration_reported_effect}
\end{align}
Each step has a direct interpretation. The first line compares reported post-treatment and pre-treatment traffic. The second line substitutes the definition of the true treatment effect. The final line separates the reported change into two components. The first component is the true effect evaluated under post-treatment calibration. The second component is the amount by which the platform revalues the website's baseline traffic because calibration changed between the two periods.

Subtracting the true effect from the reported effect gives the bias:
\begin{equation}
  \theta^{\mathrm R}-\theta
  =(\rho_1-1)\theta+(\rho_1-\rho_0)y_0.
  \label{eq:calibration_bias}
\end{equation}
The first term is the scaling error in the true effect evaluated at post-treatment calibration. The second term is the baseline-revaluation component: it captures how the change in calibration alters the reported value of baseline traffic. Because $\rho_1$ can itself change with treatment, treatment-induced calibration error can affect both terms. The baseline-revaluation component can be important even when the true effect is modest because it multiplies the entire baseline level $y_0$. A small change in calibration can therefore create a large change in reported traffic for a website with a large baseline.

Three special cases clarify the result. If $\rho_0=\rho_1=1$, reported and true effects coincide. If $\rho_0=\rho_1=\rho\neq1$, calibration is imperfect but stable, and the reported effect is simply $\rho\theta$. The baseline-revaluation component appears only when $\rho_1\neq\rho_0$. When $\rho_1>\rho_0$, this component is positive: it reinforces the measured effect $\rho_1\theta$ for a true increase and offsets it for a true decline. When $\rho_1<\rho_0$, the component is negative: it offsets the measured effect for a true increase and reinforces it for a true decline. The net bias relative to $\theta$ depends on both terms in Equation~\eqref{eq:calibration_bias}.

Because the change in $\rho_t$ has no general sign restriction, this broader mechanism can generate attenuation, amplification, or sign reversal. For a true increase, $\theta>0$, the reported effect reverses sign when
\begin{equation}
  (\rho_0-\rho_1)y_0>\rho_1\theta.
  \label{eq:calibration_positive_reversal}
\end{equation}
In words, a sufficiently large decrease in the calibration ratio outweighs the measured part of the true increase. For a true decline, $\theta<0$, the reported effect instead becomes positive when
\begin{equation}
  (\rho_1-\rho_0)y_0>\rho_1|\theta|.
  \label{eq:calibration_negative_reversal}
\end{equation}
Here, a sufficiently large increase in the calibration ratio makes reported traffic rise even though true traffic falls. These conditions also show why scale-dependent observability is a narrower case. In this two-period illustration, if calibration depends only on scale through a fixed function $\rho_t=\rho(y_t)$ and $y\rho(y)$ is strictly increasing in $y$, expected reported traffic preserves the sign of the true change, although its magnitude may differ. Treatment-induced calibration error imposes no such restriction because treatment can change calibration even at a given level of true traffic.

\subsection{What a Comparison Group Does and Does Not Solve}

We now allow calibration ratios to change for reasons unrelated to treatment, while retaining the assumption that true traffic would remain constant for both treated and control websites in the absence of treatment. A comparison group can remove calibration changes that generate the same change in reported traffic for treated and control websites. It does not automatically remove treatment-induced calibration changes. Let $\rho_{T,t}$ and $\rho_{C,t}$ denote the calibration ratios for the treated and control websites, respectively, and let $\theta$ denote the treatment effect on true traffic at the treated website. The control website is unaffected by treatment. The reported difference-in-differences estimand is
\begin{equation}
  \theta^{\mathrm{R,DiD}}
  = \left[\rho_{T,1}(y_{T,0}+\theta)
    -\rho_{T,0}y_{T,0}\right]
  - \left[\rho_{C,1}y_{C,0}
    -\rho_{C,0}y_{C,0}\right].
  \label{eq:calibration_did}
\end{equation}
If control calibration is stable, so that $\rho_{C,1}=\rho_{C,0}$, the second bracket is zero, but the treated website's calibration change remains. More generally, the calibration-change components cancel only when $(\rho_{T,1}-\rho_{T,0})y_{T,0}=(\rho_{C,1}-\rho_{C,0})y_{C,0}$. Even then, the true treatment effect is scaled by $\rho_{T,1}$. An additive calendar-time fixed effect can absorb a common additive revision. In the multiplicative representation here, however, even a common change in the calibration ratio produces different level changes when websites have different baseline traffic, so it generally does not cancel in a levels specification. A treatment-specific calibration change likewise remains because treated traffic acquires a different composition, scale, or route.

\subsection{Interpretation and Scope}

Several settings in Table~\ref{tab:traffic_peer_reviewed} illustrate why the mechanism is plausible. Product-discovery features can move small websites from sparse to readily observed traffic and can change the composition of their visitors \citep{cao_koning_nanda_2024,wright_2026}. Privacy regulation can change both user consent and the inputs available for measurement; \citet{miller2024impact} explicitly discusses this possibility and compares Similarweb estimates with an independent benchmark. New traffic sources can also create attribution gaps. For example, \citet{zhou2026llms} documents missing periods in Semrush's platform-specific AI-referral data. These examples do not establish that a particular platform made a particular calibration error. Instead, they show that treatments can change the conditions under which calibration is performed.

This mechanism is broader than observability that varies only with website scale. In the scale-based case, measurement changes because a website becomes larger or smaller. The broader mechanism also allows treatment to change audience composition, geography, device use, privacy choices, referral visibility, or other inputs used by the prediction model. It is also distinct from within-source smoothing and cross-source leakage. Those mechanisms describe how traffic observations may be transformed over time or across sources. Treatment-induced calibration error instead describes a failure in the expansion from observed signals to estimated total traffic. The mechanisms can coexist, but none is required for the others to operate.

Identification of the causal effect on true traffic therefore requires both a valid behavioral research design and a measurement process that preserves its identifying variation. In the regression framework of Equation~\eqref{eq:general_nonclassical_bias}, the additional measurement condition is $\operatorname{Cov}(X_{i,r,t},u_{i,s,t})=0$. In a two-period difference-in-differences design, this requires the expected change in measurement error to be the same for treated and control websites. Hence, similar evolution of the true-to-reported mapping in the absence of treatment is not sufficient: even a common, time-invariant multiplicative distortion can rescale the treatment effect. Treatment-induced changes in the representativeness of the platform's inputs can violate the measurement condition even when the underlying behavioral design remains valid.

\FloatBarrier
\clearpage

\section{Simulation Design and Implementation}
\label{appsec:simulation_details}

\setcounter{equation}{0}
\setcounter{table}{0}

This appendix documents the data-generating processes, platform-measurement transformations, and estimates underlying Section~\ref{sec:simulation}. The implementation uses fixed random-number seeds and holds both the focal- and auxiliary-source data-generating processes, including their realized shocks and treatment responses, constant across all three simulation conditions. As a favorable benchmark, it abstracts from upstream signal and calibration error and applies the two transformations directly to the realized traffic paths. Only the platform's loading on the auxiliary source changes. Thus, differences across the attenuation, amplification, and sign-reversal results arise entirely from the platform-measurement transformations described below.

\subsection{Design and Implementation}
\label{appssec:simulation_design}

We simulate inbound traffic attributed to a focal source for two websites over $120$ days, indexed by $t\in\{0,\ldots,119\}$. The dates run from January~1 through April~29, 2024. Website $i=1$ is treated and website $i=0$ is the control, so $T_i=\mathds{1}\{i=1\}$. The treatment begins at $t=60$ (March~1), leaving $60$ pre-treatment and $60$ post-treatment observations for each website. Defining $P_t=\mathds{1}\{t\geq60\}$, the interaction $D_{it}=T_iP_t$ equals one only for the treated website during the post-treatment period. True focal-source traffic follows
\begin{equation}
  y_{i,s,t}
  = \alpha_i + \kappa t
  + A\sin\!\left(\frac{2\pi t}{14}\right)
  + \theta_sD_{it}
  + \varepsilon_{i,s,t},
  \qquad
  \varepsilon_{i,s,t}\stackrel{\mathrm{iid}}{\sim}
  \mathcal{N}(0,\sigma_{\varepsilon}^{2}).
  \label{eq:simulation_true_dgp}
\end{equation}
The website effect $\alpha_i$ captures time-invariant differences in traffic levels, $\kappa$ produces a common linear trend, and the sine term introduces a common two-week cycle with amplitude $A$. We set $\alpha_0=1{,}000$, $\alpha_1=1{,}030$, $\kappa=2$, $A=35$, $\sigma_{\varepsilon}=15$, and $\theta_s=-120$. Thus, treatment reduces expected focal-source traffic at the treated website by $120$ units per day and leaves true traffic at the control website unchanged. These choices produce parallel trends in expectation while retaining idiosyncratic daily variation. The implementation uses seed $42$ for the focal-source shocks.

The focal source's no-event conditional mean is
\[
  y_{i,s,t}^{0}
  =\alpha_i+\kappa t+A\sin(2\pi t/14).
\]
This quantity is the simulation counterpart of the counterfactual conditional mean introduced in Section~\ref{sec:implications}. It also provides the regular pattern around which the platform smooths recent focal-source traffic.

We first construct the within-source component. It predicts current focal-source traffic using the previous $L=60$ deviations from the no-treatment conditional mean:
\begin{equation}
  \widetilde{y}^{\mathrm W}_{i,s,t}
  =y^{0}_{i,s,t}
  +\frac{1}{L}\sum_{\ell=1}^{L}
  \left(y_{i,s,t-\ell}-y^{0}_{i,s,t-\ell}\right),
  \qquad L=60.
  \label{eq:simulation_within_component}
\end{equation}
The conditional mean carries the website level, trend, and cycle into the prediction, while the trailing average carries forward recent deviations from that pattern. For negative time indices needed at the start of the sample, the code generates $60$ independent pre-sample shocks from $\mathcal{N}(0,\sigma_{\varepsilon}^{2})$ using seed $20{,}260{,}902$. The current observation never enters its own forecast. Consequently, the treatment has no effect on the within-source component on its first day and enters the component gradually as post-treatment observations replace pre-treatment observations in the window. Even before treatment, the within-source component differs from true traffic because it averages lagged shocks rather than containing the current shock.

The cross-source component uses contemporaneous traffic from auxiliary source $q$ serving the same website. Auxiliary-source traffic follows
\begin{equation}
  y_{i,q,t}
  =\alpha_{i,q}+\kappa t
  +A\sin\!\left(\frac{2\pi t}{14}\right)
  +\theta_{q}D_{it}
  +\varepsilon_{i,q,t},
  \qquad
  \varepsilon_{i,q,t}\sim\mathcal{N}(0,\sigma_{\varepsilon}^{2}).
  \label{eq:simulation_auxiliary_dgp}
\end{equation}
We set $\alpha_{0,q}=700$, $\alpha_{1,q}=720$, and $\theta_q=-120$. The auxiliary source therefore has the same treatment response, trend, cycle, and innovation variance as the focal source. Its shocks are $i.i.d.$ across websites and dates and are independent of all focal-source and pre-sample shocks. The code uses seed $20{,}260{,}901$ for these shocks and reuses the same realized auxiliary-source path in every condition. Following Equation~\eqref{eq:cross_pooling}, let $b_j$ denote the condition-specific loading that the platform places on deviations of auxiliary-source traffic from its no-event conditional mean,
\[
  y^0_{i,q,t}
  =\alpha_{i,q}+\kappa t+A\sin(2\pi t/14).
\]
The cross-source component is
\begin{equation}
  \widetilde{y}^{\mathrm C(j)}_{i,s,t}
  =y^0_{i,s,t}
  +b_j\left(y_{i,q,t}-y^0_{i,q,t}\right).
  \label{eq:simulation_cross_component}
\end{equation}
Centering the auxiliary series around $y^0_{i,q,t}$ aligns the expected untreated path of the cross-source component with the focal source's no-event conditional mean. The component nevertheless retains the auxiliary source's daily shocks and imports the treatment response $b_j\theta_q$. Reported and true traffic can therefore differ before treatment even though the simulation adds no upstream signal or calibration error.

Let $\omega=0.75$ denote the share of the modeled component assigned to cross-source information in every condition. The complete modeled component is
\begin{equation}
  \widetilde{y}^{(j)}_{i,s,t}
  =(1-\omega)\widetilde{y}^{\mathrm W}_{i,s,t}
  +\omega\widetilde{y}^{\mathrm C(j)}_{i,s,t}.
  \label{eq:simulation_nested_component}
\end{equation}
The interior value of $\omega$ combines the two mechanisms while keeping their relative shares fixed across conditions. Under the favorable signal benchmark described above, the platform places weight $\lambda=0.85$ on this modeled component and weight $1-\lambda=0.15$ on the current realized focal-source path:
\begin{equation}
  \widehat{y}^{(j)}_{i,s,t}
  =(1-\lambda)y_{i,s,t}
  +\lambda\widetilde{y}^{(j)}_{i,s,t}.
  \label{eq:simulation_reported_outcome}
\end{equation}
Equations~\eqref{eq:simulation_within_component}--\eqref{eq:simulation_reported_outcome} generate the entire reported path. Accordingly, every daily difference between reported and true traffic is generated by within-source smoothing, cross-source leakage, or both; the simulation deliberately adds no upstream signal or calibration error.

The mechanism parameters also give a simple prediction for the event response. For the treated website, let $h=0$ on the first event day and define $g_h=\min\{h,L\}/L$. The within-source component has expected response $g_h\theta_s$ because exactly $\min\{h,L\}$ event observations appear in its trailing window. The expected response in reported traffic is $m_{j,h}\theta_s$, where
\begin{equation}
  m_{j,h}
  =1-\lambda
  +\lambda\left[
    (1-\omega)g_h
    +\omega b_j\frac{\theta_q}{\theta_s}
  \right].
  \label{eq:simulation_response_multiplier}
\end{equation}
The first treatment day has $g_0=0$, so a smoothing-only reported series initially preserves only the $1-\lambda$ share of the true decline. The within-source response then grows as the window updates. Over the $60$ post-treatment observations, $h\in\{0,\ldots,59\}$ and $\overline g=29.5/60=0.4917$. Let $\overline m_j$ denote Equation~\eqref{eq:simulation_response_multiplier} evaluated at $\overline g$. Table~\ref{tab:simulation_mechanism_calibrations} reports the exact code calibrations and their implied average responses.

\begin{table}[H]
  \centering
  \small
  \caption{Mechanism-Based Simulation Calibrations}
  \label{tab:simulation_mechanism_calibrations}
  \begin{threeparttable}
  \begin{tabular}{llrrrr}
    \toprule
    Condition & Mechanism & $b_j$ & $\theta_q$ & $\overline m_j$ & $\overline m_j\theta_s$ \\
    \midrule
    Attenuation & Both mechanisms & $1/2$ & $-120$ & 0.573 & $-68.788$ \\
    Amplification & Both mechanisms & $11/6$ & $-120$ & 1.423 & $-170.788$ \\
    Sign reversal & Both mechanisms & $-5/6$ & $-120$ & $-0.277$ & 33.213 \\
    \bottomrule
  \end{tabular}
  \begin{tablenotes}[flushleft]
    \footnotesize
    \item Notes: The focal- and auxiliary-source effects are fixed at $\theta_s=\theta_q=-120$, the modeled-component weight is $\lambda=0.85$, the cross-source share is $\omega=0.75$, and the within-source window is $L=60$. Only the platform's cross-source loading $b_j$ varies across conditions. The multiplier $\overline m_j$ describes the expected event response averaged across the 60 post-treatment days.
  \end{tablenotes}
  \end{threeparttable}
\end{table}

Every condition assigns 25 percent of the modeled component to the trailing focal-source forecast and 75 percent to the cross-source component. In the attenuation condition, $b_j=1/2$, so the fixed auxiliary-source decline of 120 units contributes an imported response of $-60$. In the amplification condition, $b_j=11/6$, so the same auxiliary-source decline contributes an imported response of $-220$. In the sign-reversal condition, $b_j=-5/6$, so the fixed auxiliary-source decline contributes an imported response of $100$. Both mechanisms therefore enter every condition, but only the platform's loading on the unchanged auxiliary-source path varies. The resulting expected reported responses are $-68.788$, $-170.788$, and 33.213, respectively.

Let $z_{it}$ denote the outcome used in the regression. We set $z_{it}=y_{i,s,t}$ for true traffic and $z_{it}=\widehat y_{i,s,t}^{(j)}$ for the reported outcome in scenario $j$. For each outcome, we estimate
\begin{equation}
  z_{it}
  = \alpha_i^{z}+\eta_t^{z}+\theta^{z}D_{it}+\nu_{it}^{z},
  \label{eq:appendix_simulation_did_regression}
\end{equation}
The terms $\alpha_i^{z}$ and $\eta_t^{z}$ are website and day fixed effects. The code estimates this equation by OLS; each regression contains $240$ observations. Standard errors use the known Gaussian simulation DGP, accounting for temporal dependence induced by the 60-day trailing window and randomness in its pre-sample initialization. Two-sided normal tests evaluate whether each outcome's expected DiD coefficient over the simulation window is zero. Let $\widehat\theta$ denote the coefficient from the regression using $y_{i,s,t}$ as the outcome, let $\widehat\theta^{\mathrm R}_j$ denote the coefficient using $\widehat y_{i,s,t}^{(j)}$, let $\widehat\theta^{\mathrm W}$ denote the coefficient using $\widetilde y^{\mathrm W}_{i,s,t}$, and let $\widehat\theta^{\mathrm C}_j$ denote the coefficient using $\widetilde y^{\mathrm C(j)}_{i,s,t}$. Because the OLS coefficient is linear in the outcome, the finite-sample point estimates satisfy
\begin{equation}
  \widehat\theta^{\mathrm R}_j
  =(1-\lambda)\widehat\theta
  +\lambda\left[
    (1-\omega)\widehat\theta^{\mathrm W}
    +\omega\widehat\theta^{\mathrm C}_j
  \right].
  \label{eq:simulation_finite_sample_relation}
\end{equation}
Equation~\eqref{eq:simulation_finite_sample_relation} links each reported-outcome estimate exactly to the realized mechanism-generated components. Sampling variation makes the realized estimates differ slightly from the expected responses in Table~\ref{tab:simulation_mechanism_calibrations}. The calibration checks imposed by the code require the attenuation estimate to lie between the true estimate and zero, the amplification estimate to be more negative than the true estimate, and the sign-reversal estimate to be positive.

\subsection{Attenuation}

Attenuation means that reported traffic moves in the same direction as true traffic with a smaller average response. In this calibration, the modeled path assigns 25 percent weight to the within-source component and 75 percent to the cross-source component. On the first treatment day, the trailing forecast still contains 60 pre-treatment deviations and has no treatment response, while $b_j=1/2$ converts the fixed auxiliary-source decline of 120 units into an imported decline of 60 units. The expected first-day reported response is therefore $0.15(-120)+0.85[0.25(0)+0.75(-60)]=-56.250$. As post-treatment observations enter the forecast window, the reported decline becomes larger. Its expected average response over the full post-treatment period is $-68.788$.

Table~\ref{tab:simulation_attenuation} reports the estimates. The true-outcome coefficient is $-115.753$, close to the imposed effect of $-120$. The reported-outcome coefficient is $-68.203$, close to the expected average response generated by the two mechanisms. Its magnitude is 59 percent of the true-outcome estimate. This result gives the attenuation argument from the preceding section a concrete time-series interpretation. A platform that carries recent history into its current estimate while also loading only partially on an auxiliary-source response can reduce the average contrast used by a short-window event study.

\begin{table}[H]
  \centering
  \small
  \caption{Difference-in-Differences Estimates under Attenuation}
  \label{tab:simulation_attenuation}
  \begin{threeparttable}
  \begin{tabular}{lccccc}
    \toprule
    \multicolumn{1}{c}{Outcome} & Multiplier & Expected response & Estimate & SE & $p$-value \\
    \midrule
    True traffic & 1.000 & $-120.000$ & $-115.753$ & $(3.873)$ & $<0.001$ \\
    Reported traffic & 0.573 & $-68.788$ & $-68.203$ & $(1.558)$ & $<0.001$ \\
    \bottomrule
  \end{tabular}
  \begin{tablenotes}[flushleft]
    \footnotesize
    \item Notes: Exact Gaussian DGP-based standard errors appear in parentheses. Both regressions contain 240 observations, website fixed effects, and day fixed effects. The reported outcome uses $\lambda=0.85$, $\omega=0.75$, the 60-day within-source component, the fixed auxiliary-source response $\theta_q=-120$, and $b_j=1/2$. The multiplier and response are averages across the 60 post-treatment days.
  \end{tablenotes}
  \end{threeparttable}
\end{table}

\subsection{Amplification}

Amplification means that reported traffic moves in the same direction as true traffic with a larger average response. The modeled component assigns 25 percent weight to the within-source forecast and 75 percent to the cross-source component. Both focal- and auxiliary-source traffic fall by 120 units, but the platform loading $b_j=11/6$ converts the auxiliary decline into an imported response of $-220$. Because the within-source component has an average expected response of $0.4917(-120)=-59$, the expected reported response is $0.15(-120)+0.85[0.25(-59)+0.75(-220)]=-170.788$.

Table~\ref{tab:simulation_amplification} shows the resulting estimates. The reported-outcome coefficient is $-170.275$, compared with $-115.753$ for true traffic. The realized estimate is close to the expected response and is about 47 percent larger in magnitude than the true-outcome estimate. This result matches the cross-source logic developed in Equation~\eqref{eq:cross_pooling_response}. Although within-source smoothing pulls the response toward zero, the platform's loading on the auxiliary-source decline is large enough to produce net amplification.

\begin{table}[H]
  \centering
  \small
  \caption{Difference-in-Differences Estimates under Amplification}
  \label{tab:simulation_amplification}
  \begin{threeparttable}
  \begin{tabular}{lccccc}
    \toprule
    \multicolumn{1}{c}{Outcome} & Multiplier & Expected response & Estimate & SE & $p$-value \\
    \midrule
    True traffic & 1.000 & $-120.000$ & $-115.753$ & $(3.873)$ & $<0.001$ \\
    Reported traffic & 1.423 & $-170.788$ & $-170.275$ & $(4.625)$ & $<0.001$ \\
    \bottomrule
  \end{tabular}
  \begin{tablenotes}[flushleft]
    \footnotesize
    \item Notes: Exact Gaussian DGP-based standard errors appear in parentheses. Both regressions contain 240 observations, website fixed effects, and day fixed effects. The reported outcome uses $\lambda=0.85$, $\omega=0.75$, the 60-day within-source component, the fixed auxiliary-source response $\theta_q=-120$, and $b_j=11/6$.
  \end{tablenotes}
  \end{threeparttable}
\end{table}

\subsection{Sign Reversal}

Sign reversal occurs when reported and true traffic respond in opposite directions. The modeled component places 25 percent weight on the within-source forecast and 75 percent weight on the cross-source component. Auxiliary-source traffic falls by 120 units, as it does in the other conditions, but the negative platform loading $b_j=-5/6$ converts this decline into an imported response of 100. Averaging Equation~\eqref{eq:simulation_response_multiplier} across the post-event period gives
\begin{equation}
  \overline m_j\theta_s
  =\left\{
    0.15+0.85\left[
      0.25\left(\frac{59}{120}\right)
      +0.75\left(-\frac{5}{6}\right)
    \right]
  \right\}(-120)
  =33.2125
  \approx33.213.
  \label{eq:simulation_reversal_calculation}
\end{equation}
The negatively loaded auxiliary-source decline produces a positive imported response large enough to offset the retained focal-source decline and the gradual decline in the within-source forecast. Reported traffic is therefore expected to increase even though true traffic from both sources decreases.

Table~\ref{tab:simulation_reversal} confirms the reversal in the regression. The reported-outcome estimate is 33.869, close to the expected increase of 33.213, while the true-outcome estimate is $-115.753$. The estimate using reported traffic indicates a statistically significant increase, whereas the estimate using true traffic indicates a statistically significant decline. This case illustrates the most consequential version of the nonclassical measurement error described in Equation~\eqref{eq:general_nonclassical_bias}.

\begin{table}[H]
  \centering
  \small
  \caption{Difference-in-Differences Estimates under Sign Reversal}
  \label{tab:simulation_reversal}
  \begin{threeparttable}
  \begin{tabular}{lccccc}
    \toprule
    \multicolumn{1}{c}{Outcome} & Multiplier & Expected response & Estimate & SE & $p$-value \\
    \midrule
    True traffic & 1.000 & $-120.000$ & $-115.753$ & $(3.873)$ & $<0.001$ \\
    Reported traffic & $-0.277$ & 33.213 & 33.869 & $(2.266)$ & $<0.001$ \\
    \bottomrule
  \end{tabular}
  \begin{tablenotes}[flushleft]
    \footnotesize
    \item Notes: Exact Gaussian DGP-based standard errors appear in parentheses. Both regressions contain 240 observations, website fixed effects, and day fixed effects. The reported outcome uses $\lambda=0.85$, $\omega=0.75$, the 60-day within-source component, the fixed auxiliary-source response $\theta_q=-120$, and $b_j=-5/6$. The multiplier and response are averages across the 60 post-treatment days.
  \end{tablenotes}
  \end{threeparttable}
\end{table}

\FloatBarrier
\end{appendices}

\end{document}